\documentclass[%
preprint,
superscriptaddress,
 amsmath,amssymb,
 aps,
prl,
]{revtex4-1}

\usepackage{graphicx}
\usepackage{dcolumn}
\usepackage{bm}
\usepackage{color}
\usepackage{siunitx}
\usepackage{textcomp,mathcomp}
\usepackage{braket}
\usepackage{soul}

\begin{document}


\title{The origin of enhanced interfacial perpendicular magnetic anisotropy in LiF-inserted Fe/MgO interface}

\author{Shoya Sakamoto}
\email{shoya.sakamoto@issp.u-tokyo.ac.jp}
\affiliation{The Institute for Solid State Physics, The University of Tokyo, Kashiwa, Chiba 277-8581, Japan}

\author{Takayuki Nozaki}%
\affiliation{National Institute of Advanced Industrial Science and Technology(AIST), Research Center for Emerging Computing Technologies, Tsukuba, Ibaraki 305-8568, Japan}

\author{Shinji Yuasa}
\affiliation{National Institute of Advanced Industrial Science and Technology(AIST), Research Center for Emerging Computing Technologies, Tsukuba, Ibaraki 305-8568, Japan}

\author{Kenta Amemiya}%
\affiliation{Institute of Materials Structure Science, KEK, Tsukuba 305-0801, Japan}

\author{Shinji Miwa}
\email{miwa@issp.u-tokyo.ac.jp}
\affiliation{The Institute for Solid State Physics, The University of Tokyo, Kashiwa, Chiba 277-8581, Japan}
\affiliation{CREST, Japan Science and Technology Agency (JST), 4-1-8 Honcho Kawaguchi, Saitama 332-0012, Japan}
\affiliation{Trans-scale Quantum Science Institute, The University of Tokyo, Bunkyo, Tokyo 113-0033, Japan}

\date{\today}

\begin{abstract}
The Fe/MgO interface is an essential ingredient in spintronics as it shows giant tunneling magnetoresistance and strong perpendicular magnetic anisotropy (PMA). A recent study demonstrated that the insertion of an ultra-thin LiF layer between the Fe and MgO layers enhances PMA significantly.
In this study, we perform x-ray magnetic circular dichroism measurements on Fe/LiF/MgO multilayers to reveal the origin of the PMA enhancement. 
We find that the LiF insertion increases the orbital-magnetic-moment anisotropy and thus the magnetic anisotropy energy.
We attribute the origin of this orbital-magnetic-moment-anisotropy enhancement to the stronger electron localization and electron-electron correlation or the better interface quality with fewer defects.
\end{abstract}

\pacs{Valid PACS appear here}
\maketitle


\section{Introduction}

Fe(CoB)/MgO-based interfaces are indispensable in spintronics applications such as magnetoresistive random access memories (MRAM) because they exhibit giant tunneling magnetoresistance (TMR) effect \cite{Butler:2001aa, Mathon:2001aa, Yuasa:2004aa, Parkin:2004aa} and strong interfacial perpendicular magnetic anisotropy (PMA) \cite{Yakata:2009aa,Ikeda:2010aa,Yang:2011aa,Koo:2013aa, Lambert:2013aa,Dieny:2017aa} simultaneously. 
Strong PMA is essential to reduce the size of magnetic cells with thermal stability maintained, and strengthening PMA is one of the major challenges in MRAM applications.

The strong PMA itself may be obtained by fabricating thin films with bulk PMA, such as $L1_{0}$-ordered (Fe,Co)(Pt,Pd) alloys \cite{Lairson:1993wr, Visokay:1995up, Watanabe:1996wa, Gehanno:1997wg, Sakamoto:2017wp}. However, those materials cannot simply be incorporated because large TMR also needs to be achieved \cite{Yoshikawa:2008tb, Kim:2008wk}.
In this regard, it is practical to modify Fe/MgO-based systems as one can expect an increase in PMA without deteriorating TMR.
There have been many studies along this line; for example, heavy-metal-layer insertion \cite{Worledge:2011tk,Liu:2012uf,Pai:2014ws,Skowroifmmode:2015tl} or heavy-metal doping \cite{Nozaki:2017vs,Nozaki:2018vz} were employed to bestow PMA through strong spin-orbit interaction of a heavy-metal element.

Modifying or replacing a barrier layer may also enhance PMA. Among many candidate materials, LiF is promising because it has the same NaCl-type structure as MgO and has excellent lattice matching with Fe ($a_{\rm Fe}=2.87$ \AA, $a_{\rm LiF}/\sqrt{2}=2.84$ \AA, $a_{\rm MgO}/\sqrt{2} = 2.98$ \AA).
Moreover, a previous theoretical study predicted that the Fe/LiF/Fe trilayer should exhibit a large TMR of 2400\% \cite{Vlaic:2016us}, although reported experimental TMR ratios were no more than 20 \% at room temperature because of the poor crystallinity of LiF \cite{Xue:2016ti,Narayananellore:2018uf}.

Recently, Nozaki {\it et al}. demonstrated that the insertion of an ultra-thin LiF layer between Fe and MgO remarkably enhances the interfacial PMA while maintaining or even increasing TMR \cite{Nozaki:2022ug}.
Despite this promising discovery, the origin of the enhancement remains to be clarified, and such clarification is critical for further PMA improvement.

In the present study, we perform x-ray magnetic circular dichroism (XMCD) measurements on Fe/LiF/MgO multilayers to reveal the origin of the PMA enhancement. 
XMCD is one of the most suitable methods for such a purpose \cite{Gambardella:2003wf, Okabayashi:2014tr, Kanai:2014wy, Miwa:2017ud, Sakamoto:2021wq} because XMCD can measure orbital-magnetic-moment anisotropy (OMA) and an intra-atomic magnetic dipole operator term, both of which determine PMA \cite{Bruno:1989aa,Laan:1998ty,Stohr:1999aa}.
We find that the LiF insertion increases the OMA and thus the PMA energy. We attribute the strengthened OMA to improved interface quality or a reinforced electron localization and orbital polarization at the Fe/LiF interface.

\section{Methods}
\begin{figure}
\begin{center}
\includegraphics[width=8.3 cm]{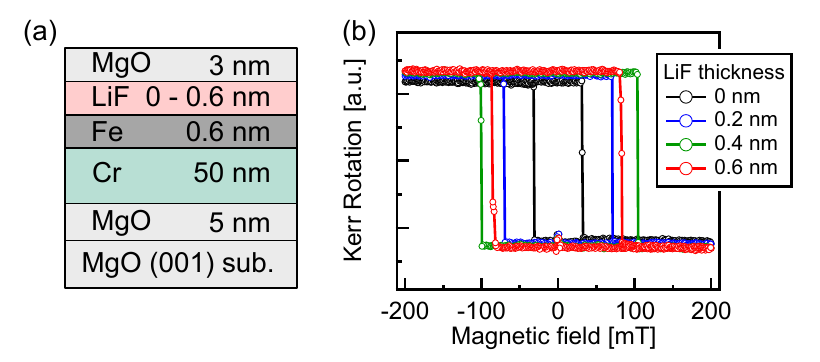}
\caption{(a) Schematic sample structure. (b) Out-of-plane magnetic hysteresis loops measured using magneto-optical Kerr effect (MOKE).}
\label{Fig1}
\end{center}
\end{figure}

The Fe/LiF/MgO heterostructures were grown on single-crystalline MgO(001) substrates by molecular beam epitaxy \cite{Nozaki:2022ug}.
The sample structure (Fig. \ref{Fig1}(a)) was MgO(001) substrate/MgO (5 nm)/Cr (50 nm)/Fe (0.6 nm)/LiF (0 - 0.6 nm)/MgO (3 nm).
A 5-nm-thick MgO seed layer and a 50-nm-thick Cr buffer layer were deposited at \SI{200}{\celsius} and annealed at \SI{800}{\celsius}.
A 0.6-nm-thick Fe(001) layer was then grown at \SI{150}{\celsius} and annealed at \SI{250}{\celsius}.
A LiF layer with thicknesses of 0, 0.2, 0.4, 0.6 nm and a 3-nm-thick MgO capping layer were grown at room temperature and annealed at \SI{250}{\celsius}.

X-ray absorption spectroscopy (XAS) and XMCD measurements were performed at the beamline BL-16A in the Photon Factory \cite{Amemiya:2013wl}.
The measurement temperature was room temperature, and magnetic fields ($H$) of 5 T were applied parallel to incident x rays.
The samples were placed such that the surface normal direction becomes parallel to the incident x rays (normal incidence) or tilted by 70 degrees (grazing incidence).
The XAS and XMCD spectra were recorded in the total-electron-yield mode. XMCD signals were obtained by reversing the helicity of x rays with 10-Hz frequency at each photon energy under a fixed magnetic field \cite{Amemiya:2013wl}, which allowed us to eliminate time-dependent background signals. The measurements were repeated with the opposite magnetic field direction to minimize artifacts.

\section{Results}

Figure \ref{Fig1}(b) shows an out-of-plane magnetic hysteresis loop measured using the magneto-optical Kerr effect (MOKE). 
The hysteresis loops have a perfectly square shape, indicating that the Fe/LiF/MgO multilayers exhibit PMA.
The coercive field increases with the LiF thickness up to 0.4 nm but decreases slightly when the LiF layer becomes 0.6-nm thick. This behavior is consistent with a previous study \cite{Nozaki:2022ug}, suggesting that the PMA energy increases upon the LiF insertion.

\begin{figure}
\begin{center}
\includegraphics[width=8.3 cm]{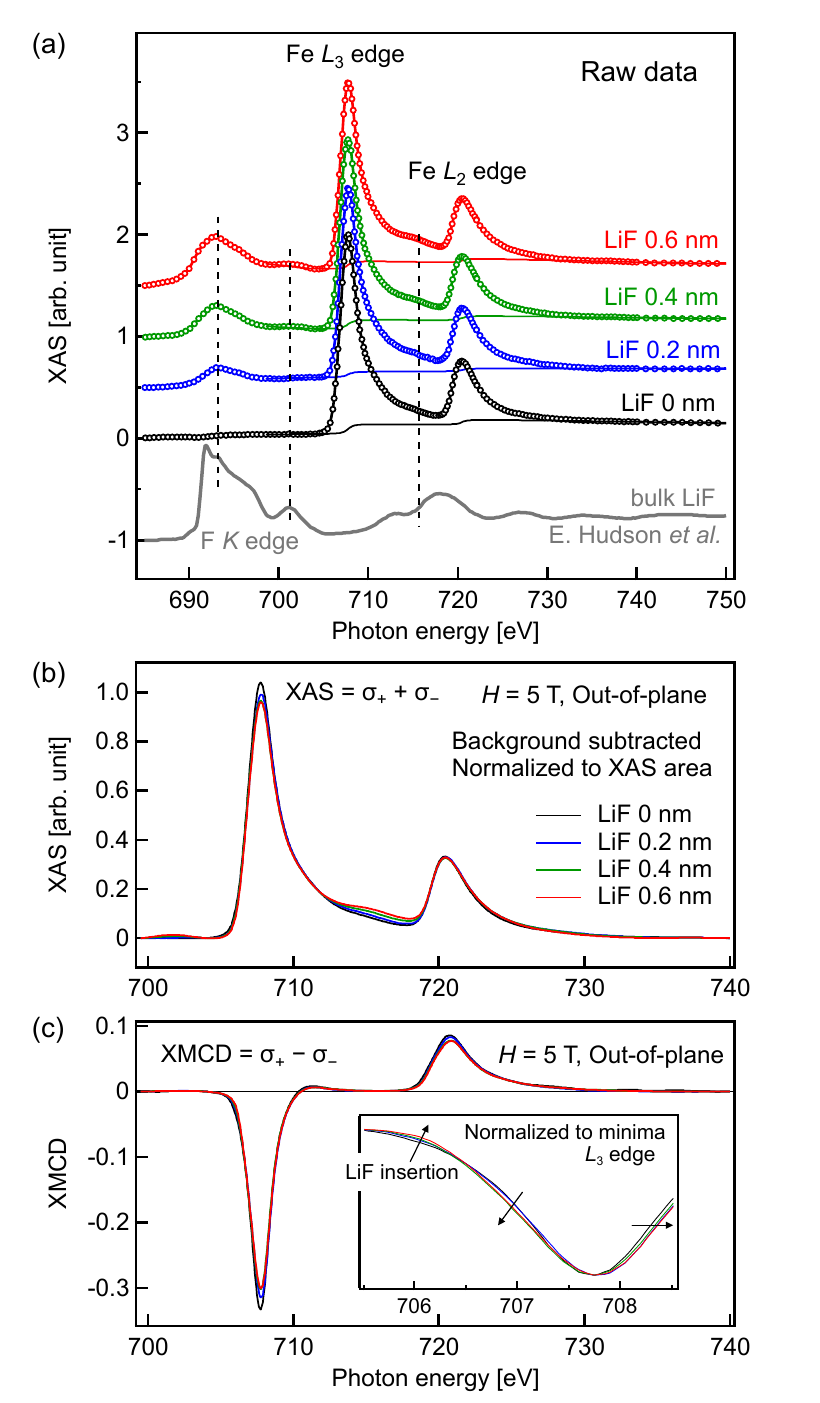}
\caption{(a) XAS spectra of the Fe/LiF/MgO multilayers. The background consisting of a double-step function and a linear function that bends at the Fe $L_{3}$ edge is also shown. XAS spectrum of bulk LiF at the F $K$ edge \cite{Hudson:1994vx}. (b) XAS spectra normalized to their area after the background subtraction and (c) corresponding XMCD spectra. The magnetic field was applied in the out-of-plane direction.}
\label{Fig2}
\end{center}
\end{figure}

Figure \ref{Fig2}(a) shows typical Fe $L_{2,3}$-edge XAS spectra of the Fe/LiF/MgO multilayers.
As the LiF thickness increases, peaks emerge and develop around 693, 701, and 715 eV, as indicated by dashed vertical lines in Fig. \ref{Fig2}(a).
These peaks originate from F K-edge absorption in the LiF layer, and the spectral line shapes are indeed similar to that of the F $K$-edge XAS spectrum of the bulk LiF single crystal \cite{Hudson:1994vx}, which is displayed at the bottom of Fig. \ref{Fig2}(a). 
Furthermore, the observed F K-edge XAS spectra look distinct from those of magnesium fluorides \cite{Oizumi:1985us} and iron fluorides \cite{Vinogradov:2005uc}, indicating the absence of interlayer mixing between the Fe, LiF, and MgO layers.
Despite the spectral overlap between F $K$-edge and Fe $L_{2,3}$-edge XAS spectra, we subtracted backgrounds in a rather standard way \cite{Sakamoto:2022vt}; we subtracted the summation of a double step function and a linear function that bends at the Fe $L_{3}$ edge such that XAS intensities become zero in the pre-edge (around 700 eV) and post-edge regions (around 740 eV). The backgrounds are shown by solid curves separately for each spectrum in Fig. \ref{Fig2}(a).

Figure \ref{Fig2}(b) shows the XAS spectra with the background subtracted. 
The XAS spectrum of the LiF-free Fe/MgO multilayer looks identical to those reported in previous studies \cite{Miyokawa:2004aa,Luches:2005wl, Sakamoto:2022vt, Sakamoto:2022tc}; the XAS spectrum has a broad single peak for each of the Fe $L_{2,3}$ edges and does not exhibit multiplet features. This is also the case for the spectra with LiF insertion, except that there is a slight variation in XAS spectral line shape because of the spectral overlap between F $K$ and Fe $L_{2,3}$ edges.
The absence of multiplet features indicates that the Fe do not form an ionic bond with F but instead form very weakly hybridized Fe-F orbitals as in the case of Fe-O hybridization at the Fe/MgO interface. These results are inconsistent with previous studies on polycrystalline Fe/LiF/MgO multilayers that suggested significant interlayer mixing and the formation of Fe$^{3+}$ ions at the Fe/LiF interface \cite{Kita:1992ve}.

The XMCD spectra [Fig. \ref{Fig2}(c)] do not show F K-edge features and thus only reflect the Fe electronic structure. The absence of F signals indicates that F atoms do not have a magnetic moment as expected from the filled outer 2$p$ shells and weak Fe-F hybridization.
The spectral line shape changes slightly with increasing LiF thickness, as shown in the inset of Fig. \ref{Fig2}(c); there is a spectral intensity transfer from $\sim$706 eV to $\sim$707 eV, and the spectra shift towards higher energies. This line shape change indicates that the electronic structure of Fe/LiF differs from that of the Fe/MgO interface. The spectral shift suggests that the Fe is slightly less hole-doped in the Fe/LiF interface than in the Fe/MgO interface, possibly because F can absorb fewer electrons than O.
Considering the fact that hole doping usually strengthens the PMA of Fe/MgO \cite{Maruyama:2009vw, Niranjan:2010th, Miwa:2015vb, Nozaki:2016uj, Zhang:2017aa}, the PMA enhancement found in less hole-doped Fe/LiF indicates that the Fe-F hybridization significantly alters the interfacial electronic structure.

\begin{figure}
\begin{center}
\includegraphics[width=8.3 cm]{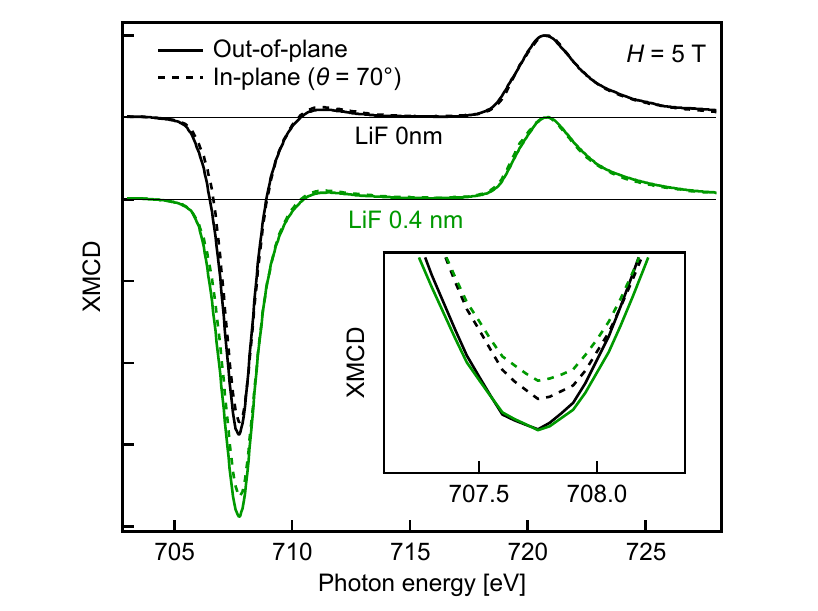}
\caption{XMCD spectra taken with out-of-plane and in-plane magnetic fields of 5 T. The spectra are normalized to the $L_{2}$ edge maximum. The inset shows the magnified view around the $L_{3}$-edge peak.}
\label{Fig3}
\end{center}
\end{figure}

To reveal the origin of PMA, we measured XMCD spectra with out-of-plane and in-plane magnetic fields.
Figure \ref{Fig3} compares the XMCD spectra measured with out-of-plane and almost in-plane (70\textdegree) magnetic fields for the samples without LiF and with 0.4-nm-thick LiF. Here, the XMCD spectra are normalized to the $L_{2}$-edge maxima.
The inset is a magnified view around the $L_{3}$-edge minima, where one can see that the XMCD intensity is stronger for the out-of-plane magnetic fields than for the in-plane magnetic fields. This intensity anisotropy becomes more prominent with the LiF insertion, meaning that the orbital magnetic moment becomes more anisotropic, as will be quantitatively discussed below. 
 
Here, we estimate the spin and orbital magnetic moments applying the XMCD sum rules \cite{Thole:1992aa, Carra:1993aa, Chen:1995aa}:  
\begin{eqnarray}
&\displaystyle m_{\rm orb}=-\frac{4}{3}\frac{S^{\rm XMCD}_{L_{3}} + S^{\rm XMCD}_{L_{2}}}{S^{\rm XAS}}n_{h},\\
&\displaystyle m_{\rm spin}^{\rm eff} = m_{\rm spin}+7m_{\rm T}=-2\frac{S^{\rm XMCD}_{L_{3}} - 2 S^{\rm XMCD}_{L_{2}}}{S^{\rm XAS}}n_{h},
\end{eqnarray}
where $m_{\rm orb}$ and $m_{\rm spin}$ are the orbital and spin magnetic moments in units of $\mu_{\rm B}$/atom, respectively, and $\mu_{\rm B}$ is the Bohr magneton. $m_{\rm T}$ is the expectation value of the magnetic dipole term $m_{\rm T} = -\langle T \rangle \mu_{\rm B}/\hbar$, with $\hbar$ being the reduced Planck constant \cite{Stohr:1999aa}.  $S^{\rm XMCD}_{L_{2,3}}$ represents the XMCD integral over the $L_{2}$ or the $L_{3}$ edges, and $S^{\rm XAS}$ represents the XAS integral over the $L_{2,3}$ edges. As the magnetic dipole term was found negligible in the Fe/MgO interface \cite{Sakamoto:2022vt}, we ignored this term hereafter.
$n_{h}$ is the number of 3$d$ holes, which is assumed to be 3.39 in the present study.

\begin{figure}
\begin{center}
\includegraphics[width=8.3 cm]{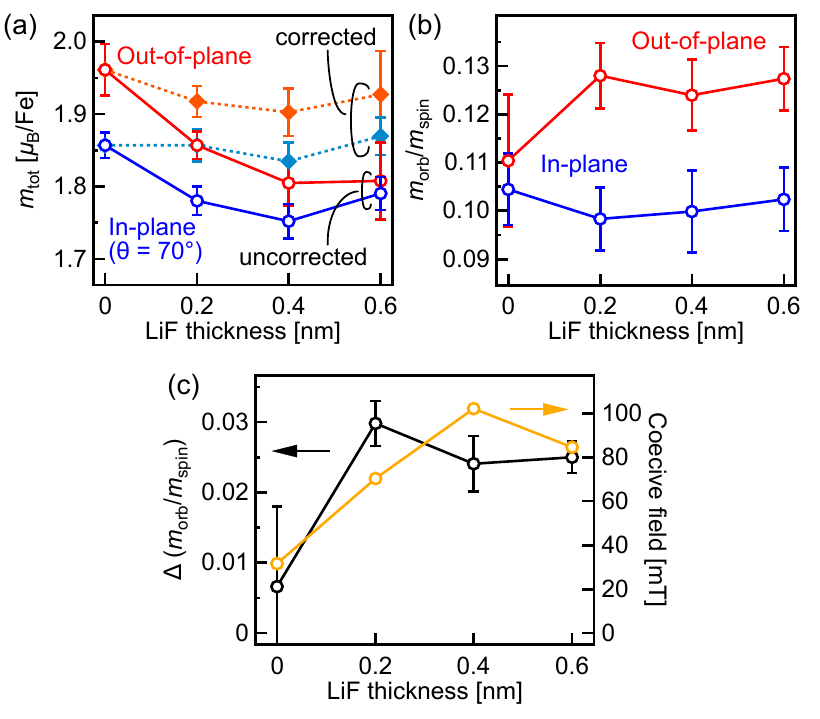}
\caption{Magnetic moments deduced using the XMCD sum rules. (a) Total magnetic moments. (b) Ratio of spin magnetic moment to orbital magnetic moment. (c) The anisotropy of orbital to spin magnetic moment ratio. The error bars represent the uncertainty in the spectral background subtraction process \cite{footnoteBGsubtraction}. The coercive field, measured by MOKE, is also plotted.}
\label{Fig4}
\end{center}
\end{figure}

The deduced total magnetic moments ($m_{\rm tot} = m_{\rm spin} + m_{\rm orb}$) are plotted in Fig. \ref{Fig4}(a) with open circles. Because the spectral overlap between F $K$-edge and Fe $L_{2,3}$-edge XAS leads to the overestimation of the XAS area ($S^{\rm XAS}$), the total magnetic moment becomes underestimated and decreases as the LiF layer becomes thick. To avoid this underestimation, we corrected the magnetic moment data by normalizing them to the XAS areas between 704 and 712 eV, where F K-edge absorption signals should almost be absent. Thus corrected total magnetic moments are plotted in Fig. \ref{Fig4}(a) with filled diamonds. Although this correction is not exact, it can be said that the LiF insertion does not drastically modulate the Fe magnetic moment. 

Despite the difficulty in quantitatively estimating the spin and orbital magnetic moments, its ratio $m_{\rm orb}/m_{\rm spin}$ can be reliably obtained since both the XAS area term and hole number term are divided out.
Figure \ref{Fig4}(b) shows deduced $m_{\rm orb}/m_{\rm spin}$ as a function of LiF thickness for out-of-plane and in-plane magnetic fields. The out-of-plane orbital magnetic moment increases as LiF thickness increases, while the in-plane orbital magnetic moment shows the opposite behavior. As a result, the anisotropy of the orbital to spin magnetic moment ratio, defined as $\Delta (m_{\rm orb}/m_{\rm spin}) = (m_{\rm orb}/m_{\rm spin})_{\theta = 0^{\circ}} - (m_{\rm orb}/m_{\rm spin})_{\theta = 70^{\circ}}$, increases with LiF thickness [Fig. \ref{Fig4}(c)]. This strengthened orbital moment anisotropy (OMA) is consistent with the PMA enhancement because the PMA energy is proportional to the OMA in the simplest approximation \cite{Bruno:1989aa}. Indeed, the coercive fields behave similarly to the OMA, as displayed in Fig. \ref{Fig4}(c).

\section{Discussion}
Here, let us discuss the possible origin of the PMA enhancement. The local magneto-crystalline anisotropy energy at site $i$, $E_{\rm MCA}^{i}$, can be written within second-order perturbation theory \cite{Wang:1993wf, Masuda:2018vg} as 
\begin{align}
E_{\rm MCA}^{i} \approx \Delta E_{\uparrow,\uparrow}^{i} + \Delta E_{\downarrow,\downarrow}^{i} + \Delta E_{\uparrow,\downarrow}^{i} + \Delta E_{\downarrow, \uparrow}^{i},\\
\Delta E_{\sigma, \sigma}^{i}  = \xi_{i}^{2} \sum_{u_{\sigma}, o_{\sigma}} \frac{|\braket{u_{\sigma}|L_{z}^{i}|o_{\sigma}}|^2 - |\braket{u_{\sigma}|L_{x}^{i}|o_{\sigma}}|^2}{\epsilon_{u_{\sigma}} - \epsilon_{o_{\sigma}}},\label{spin-conserve}\\
\Delta E_{\sigma, \sigma'}^{i}  = \xi_{i}^{2} \sum_{u_{\sigma'}, o_{\sigma}} \frac{|\braket{u_{\sigma'}|L_{x}^{i}|o_{\sigma}}|^2 - |\braket{u_{\sigma'}|L_{z}^{i}|o_{\sigma}}|^2}{\epsilon_{u_{\sigma'}} - \epsilon_{o_{\sigma}}}\label{spin-flip}.
\end{align}
Here, $\sigma$ and $\sigma'$ represent either up ($\uparrow$) or down ($\downarrow$) spin. Eqs. (\ref{spin-conserve}) and (\ref{spin-flip}) represent spin-conserving and spin-flipping scattering ($\sigma \neq \sigma'$), respectively.
$\xi_{i}$ denotes the spin-orbit coupling constant, and $\ket{o_{\sigma}}$ ($\ket{u_{\sigma}}$) represents a local occupied (unoccupied) state with spin $\sigma$ and energy $\epsilon_{o_{\sigma}}$ ($\epsilon_{u_{\sigma}}$). $L_{x,z}^{i}$ is the local angular momentum operator. 
The spin-conserving and spin-flipping terms are related to the orbital-magnetic-moment anisotropy $\Delta m_{l}$ and magnetic dipole term $m_{T}$, respectively \cite{Bruno:1989aa, Laan:1998ty}.

In the case of Fe/MgO-based interfaces, $\Delta E_{\uparrow,\uparrow}^{i}$ and $\Delta E_{\downarrow, \uparrow}^{i}$ are negligible because spin-up bands are almost fully occupied, and the interfacial PMA predominantly originates from $\Delta E_{\downarrow,\downarrow}^{i}$ and $\Delta E_{\uparrow,\downarrow}^{i}$. To be more specific, the predominant contributions were proposed to be spin-conserving scattering between $d_{xz,\downarrow}$ and $d_{yz, \downarrow}$ orbitals and spin-flipping scattering between $d_{z^{2}, \uparrow}$ and $d_{xz/yz, \downarrow}$ orbitals and between $d_{xy, \uparrow}$ and $d_{xz/yz, \downarrow}$ orbitals \cite{Masuda:2018vg}. 

In the present study, we have revealed that the LiF insertion significantly increases the OMA of Fe. This OMA enhancement should manifest as an increased spin-conserving scattering term, as mentioned above. Because the spin-conserving term was more significant for Fe/MgO with the lattice constant of $a=a_{\rm MgO}/\sqrt{2} = 2.98$ \AA\ than that with $a=a_{\rm Fe}=2.87$ \AA\ \cite{Masuda:2018vg}, the enhanced OMA cannot be solely attributed to the better lattice matching between Fe and LiF ($a_{\rm LiF}/\sqrt{2}=2.84$ \AA) but should stem from chemical-property differences between LiF and MgO. 

LiF is one of the most ionic compounds and has a much deeper valence band maximum and a much wider band gap than MgO \cite{Prada:2008uf}. This extremely ionic nature results in weaker Fe 3$d$--F 2$p$ hybridization. 
According to a previous theoretical study on an Fe monolayer sandwiched by MgO layers \cite{Lee:2013ue}, the Fe-O bond length (or the Fe 3$d_{z}^2$ and O 2$p_{z}$ hybridization strength) increases the spin-flipping term but does not significantly affect the spin-conserving term. This may call for another mechanism to explain the OMA enhancement. One possibility is that a stronger electron--electron correlation at the Fe/LiF interface enhances the OMA and PMA \cite{Sakamoto:2022vt} as the more ionic nature of LiF results in stronger interfacial electron localization with stronger electron--electron correlation.
Note that it is possible that better lattice matching just improves the interface quality and thus enhances the OMA and PMA. 

In addition to the enhanced spin-conserving term, the spin-flipping term might also contribute to PMA.
As mentioned above, a previous theoretical study \cite{Lee:2013ue} suggested that weakened Fe 3$d_{z}^2$ and O 2$p_{z}$ hybridization increases the spin-flipping term between $d_{z^{2}, \uparrow}$ and $d_{xz/yz, \downarrow}$ orbitals by lowering the energy position of the majority-spin anti-bonding 3$d_{z}^2$-2$p_{z}$ orbitals down to the Fermi level.
The same mechanism may apply to the Fe/LiF interface because LiF may be viewed as FeO with weakened Fe-O hybridization.

To further clarify the origin of the PMA and OMA enhancement, it is necessary to perform density-functional-theory calculations on the Fe/LiF interface with and without electron--electron correlation considered.

\section{Summary}
In the present work, we have performed x-ray magnetic circular dichroism measurements with out-of-plane and in-plane magnetic fields to reveal the origin of enhanced perpendicular magnetic anisotropy in Fe/LiF/MgO multilayers.
We have found that ultra-thin LiF insertion increases the orbital-magnetic-moment anisotropy of Fe and thus enhances perpendicular magnetic anisotropy.
We attribute this orbital-magnetic-moment anisotropy enhancement to the more robust interfacial electron localization and electron--electron correlation, which results from the highly ionic nature of LiF and weak Fe-F hybridization, or the better interface quality with fewer defects.

\section*{Acknowledgments}
This work was performed under the approval of the Photon Factory Program Advisory Committee (Proposal Nos. 2019S2-003 and 2020G518).
This work was partly supported by JSPS KAKENHI (Nos. JP20K15158, 22H00290, 22H04964, 22K18320), Spintronics Research Network of Japan (Spin-RNJ), and JPNP16007, commissioned by the New Energy and Industrial Technology Development Organization (NEDO), Japan.

\bibliography{../../../Bibtex/BibTex_all}

\begin{thebibliography}{57}
\bibitem [{foo()}]{footnoteBGsubtraction}%
  \BibitemOpen
  \href@noop {} {\bibinfo  {journal} {We subtracted a background from each XAS
  spectrum using about 250 different sets of parameters. Thus obtained average
  values are plotted in Fig. 4, and the standard deviations are used as error
  bars}\ }\BibitemShut {NoStop}%
\end{thebibliography}

\end{document}